\documentclass[a4paper,11pt]{article}
\usepackage[T1]{fontenc}
\usepackage{mathptmx}
\usepackage[utf8]{inputenc}
\usepackage{amsmath,amssymb,graphicx,hyperref}
\usepackage[a4paper,margin=2.5cm]{geometry}
\newcommand{\affiliation}[1]{}
\newcommand{\emailAdd}[1]{}
\long\def\abstract#1{\begin{quote}\textbf{Abstract.} #1\end{quote}}
\newcommand{\keywords}[1]{\par\noindent\textbf{Keywords: }#1\par}
\newcommand{\note}[1]{}
\usepackage{longtable,booktabs,array,calc}
\usepackage{placeins}
\usepackage{graphicx,hyperref}
\newcommand{\real}[1]{#1}
\title{Time-Resolution Function of the RADEX Neutron Source\\[0.5em]\large Determination from Geant4 Data and a Physically Motivated Gamma Parametrization}
\author{R. M. Djilkibaev\\Institute for Nuclear Research of the Russian Academy of Sciences\\7a, 60th October Anniversary Prospect, Moscow 117312, Russia\\\texttt{rmd@inr.ru}}
\affiliation{Institute for Nuclear Research of the Russian Academy of Sciences,\\7a, 60th October Anniversary Prospect, Moscow 117312, Russia}
\emailAdd{rmd@inr.ru}
\date{18 September 2026}
\begin{document}
\maketitle
\abstract{The time-resolution function is one of the key factors determining the accuracy of neutron-resonance parameter reconstruction in time-of-flight experiments. A method is proposed for determining the time-resolution function of the RADEX neutron source associated with neutron slowing down in the target, based on detailed Geant4 modeling. The neutron time spectra obtained in the Geant4 simulations are approximated by a sum of three Gamma functions with dimensionless parameters. This model is shown to describe the shape of the time distributions well over the neutron-energy range 50--300 eV. To test the applicability of the proposed approach to a source of a different design, a comparison is made with the calculated n\_TOF time-resolution function. Its shape is also shown to be well described by a three-component Gamma model.}
\keywords{neutron time-of-flight, resolution function, Geant4, Gamma distribution, RADEX, n\_TOF.}

\section{Introduction}

Time-of-flight (TOF) experiments make it possible to reconstruct
neutron-resonance parameters from the measured spectral shape; however,
the observed resonance shape is determined not only by the intrinsic
nuclear widths. It is additionally distorted by the finite proton-pulse
duration, neutron transport in the target and moderator, the geometry of
the flight path, detector response, and time binning. The
time-resolution function is therefore an integral part of the physical
model of the experiment, and an inaccurate description of it directly
shifts the reconstructed resonance energies and widths. In facilities
where neutrons are produced and moderated in an extended
target-moderator system, an important part of the time response is
associated with neutron transport before escape from the source. This
distribution is generally non-Gaussian: it is strongly asymmetric and
has a long tail corresponding to rare trajectories with many collisions
and increased residence time in matter. In a TOF experiment, the time
origin is defined by the arrival of the primary proton beam at the
target. The neutron time delay must therefore be measured from the start
of the primary proton event. In Geant4 {[}1{]}, this time scale
corresponds to GlobalTime, denoted by T below. It is measured from the
start of the primary proton event to the instant when the neutron exits
the target and includes both cascade development and the delay before
neutron birth, as well as the subsequent slowing down and transport of
the neutron in the target. In this work, the time-resolution function of
the RADEX neutron source {[}2{]} associated with neutron slowing down in
the target is studied using detailed Geant4 modeling.

\hypertarget{resolution-functions-in-previous-work}{%
\subsection{Resolution functions in previous
work}\label{resolution-functions-in-previous-work}}

Reviews of TOF spectrometers emphasize that broadening mechanisms and
intrinsic resonance shapes require a full convolution treatment, and
that the amplitude of the long tail may vary with energy {[}3{]}. For
n\_TOF {[}3{]}, the resolution function is determined by the combined
proton-pulse duration and slowing-down time in the lead target and water
moderator; its shape and energy dependence are determined primarily by
Monte Carlo calculations {[}4{]}.

Both numerical and analytical representations of the resolution function
are used in resonance analysis. SAMMY implements several
resolution-function representations, including RPI, GELINA, and n\_TOF.
The Tr108 example demonstrates the use of the dedicated n\_TOF
resolution function {[}5{]}. The RPI parametrization describes the
combined response of the neutron-producing target, moderator, and
detector system by a combination of a \ensuremath{\chi}\textsuperscript{2} component and exponential
components {[}6{]}. Because the \ensuremath{\chi}\textsuperscript{2} distribution is a special case of the
Gamma distribution, these models are naturally related to functions of
the form x\textsuperscript{n} exp(\ensuremath{-}kx) used in the present work.

For complex sources, an analytical function does not replace transport
modeling. Its purpose is to provide a compact representation of the
numerical response obtained by Monte Carlo simulation and then to use
that response in the convolution with the resonance shape. This approach
is used, in particular, for n\_TOF, where the time response is
determined by the combined effects of the proton pulse, lead target, and
water moderator {[}3,4{]}. For the very broad energy range of n\_TOF
EAR1, a flexible neural-network parametrization has also been proposed
{[}7{]}. In a limited resonance-energy region, however, a more compact
model is useful because its parameters and physical constraints can be
directly controlled.

\hypertarget{physical-origin-of-the-slowing-down-time}{%
\subsection{Physical origin of the slowing-down
time}\label{physical-origin-of-the-slowing-down-time}}

A simple cascade model of neutron slowing down in water is based on the
fact that, in the energy range considered here, neutron interaction with
\textsuperscript{1}H is dominated by elastic scattering, while the cross section varies
relatively slowly and has a characteristic scale of tens of barns
{[}8{]}. Hydrogen is an efficient moderator because the neutron and
proton masses are nearly equal. For isotropic elastic scattering on a
free proton, the neutron energy after a collision is distributed between
zero and its initial value, with a mean value approximately equal to one
half of the initial energy. Consequently, as the neutron slows down, the
time between collisions increases. The last, slowest stages make the
dominant contribution to the total delay, whereas the early stages at
higher energies contribute less. A characteristic time scale
t\textsubscript{s}(E)=0.5/\ensuremath{\sqrt{E}} \ensuremath{\mu}s, with E in eV, can be introduced; it corresponds to the
flight time over a characteristic mean free path of about 0.7 cm. The
slowing-down time in the simple cascade model is then estimated as
t\textsubscript{s}(1+1/\ensuremath{\sqrt{2}}+1/\ensuremath{\sqrt{4}}+\ldots) \ensuremath{\approx} 3.4t\textsubscript{s}. Despite its extreme simplicity, this
estimate reproduces the characteristic delay scale observed in the
Geant4 simulations. In the actual process, the number of collisions,
individual path lengths, scattering angles, and energy losses are
stochastic. The deterministic sequence therefore becomes an asymmetric
statistical distribution with a pronounced maximum and an extended tail
toward long times. A natural analytical form for such a distribution is
a Gamma-like function

\[g(x;n,k)=\frac{k(kx)^n e^{-kx}}{\Gamma(n+1)},\quad x\geq0,\qquad x=\frac{t}{t_s(E)},\qquad \langle x\rangle=\frac{n+1}{k}.\]

The power-law factor describes the accumulation of several successive
stages of the process, whereas the exponential factor suppresses the
probability of very large delays. A sum of Gamma components provides a
compact description of both the main part of the distribution and a
slower tail. Experimental and computational studies of neutron slowing
down in water directly demonstrate slowing-down-time distributions
{[}9,10{]}. The individual Gamma components are therefore treated below
as an analytical representation of the stochastic response and are not
identified directly with separate microscopic classes of neutron
trajectories.

\hypertarget{main-idea-and-objective-of-the-work}{%
\subsection{Main idea and objective of the
work}\label{main-idea-and-objective-of-the-work}}

The objective of this work is not to choose another convenient empirical
formula, but to construct the resolution function directly from detailed
transport modeling. Geant4 does not prescribe an analytical resolution
function; instead, it generates a numerical sample of individual
transport histories. For each j-th outgoing neutron, its exit energy E\textsubscript{j}
and time delay T\textsubscript{j} are stored. The dimensionless delay x\textsubscript{j}=T\textsubscript{j}/t\textsubscript{s}(E\textsubscript{j}) is
then calculated. The set of pairs (E\textsubscript{j},x\textsubscript{j}) obtained from Geant4
constitutes the input data for the subsequent statistical fit.

\[D_{\mathrm{G4}}=\{(E_j,T_j)\}_{j=1}^{N}\longrightarrow\{(E_j,x_j)\}_{j=1}^{N},\qquad x_j=\frac{T_j}{t_s(E_j)}.\]

The analytical function is introduced only at the next stage, as a model
of the conditional distribution of the delay x at a given
outgoing-neutron energy E. To distinguish the model explicitly from the
Geant4 data, its normalized probability density is denoted by
p\textsubscript{M}(x\textbar E; \ensuremath{\theta}\textsubscript{M}), where
\ensuremath{\theta}\textsubscript{M} is the set of model parameters for a mixture
containing m Gamma components.

\[p_M(x\mid E;\theta_M)=\sum_{i=1}^{m}w_i(E;\theta_M)g_i(x\mid E;\theta_M),\qquad \sum_i w_i(E;\theta_M)=1.\]

The number of components is not fixed in advance by a microscopic
interpretation, but is determined by sequential comparison of the local
W1, W2, and W3 models. Here W1, W2, and W3 denote the complete models
containing one, two, and three Gamma components, respectively; their
analytical probability densities correspond to p\textsubscript{1}, p\textsubscript{2}, and p\textsubscript{3}. In the
local fits, the parameters \ensuremath{\theta}\textsubscript{M} are determined
independently in each energy interval. Individual parameters of the
first components may be strongly correlated, so the physical conclusion
should be based primarily on the stability of the complete
resolution-function shape and on the presence of an additional time
scale, rather than on the interpretation of each mixture coefficient.

The far tail is then tested by a separate W2\_QUAD+G3 extension: a third
Gamma component, G3, is added to the fixed W2\_QUAD model. Thus, W1, W2,
W3, W2\_QUAD, and W2\_QUAD+G3 are used below as model names, whereas
p\textsubscript{M}(x\textbar E; \ensuremath{\theta}\textsubscript{M}) denotes the
corresponding normalized analytical probability density.

\hypertarget{scope-of-the-study}{%
\subsection{Scope of the study}\label{scope-of-the-study}}

The multilayer RADEX2 target is the principal object of the present
study. The monolithic RADEX1 geometry, containing the same total
thicknesses of tungsten and water, is used as a controlled reference
configuration for examining the sensitivity of the far time tail to the
internal arrangement of the target materials. The study: (i) examines
the shape of the RADEX2 Geant4 time distributions over 50--300 eV; (ii)
determines, among the models considered, the number of Gamma components
required to describe the distributions; (iii) constructs a smooth
energy-dependent model of the main RADEX2 shape; (iv) quantitatively
investigates the far time tail and compares it with the RADEX1 control
geometry; and (v) compares the result with the calculated n\_TOF
resolution function from the SAMMY Tr108 example {[}4,5{]}. The effect
of uncertainty in the resolution function on the reconstruction of
resonance parameters is discussed as an important systematic consequence
and will be the subject of a separate study.

\section{Geant4 transport modeling of RADEX}
Determination of the resolution function requires not only the source
geometry but also the statistics of actual neutron transport histories.
Therefore, the primary object of the study is not an analytical function
but the time spectrum obtained from Geant4. The simulation was performed
with Geant4 11.4 patch 01 (13 March 2026) in multithreaded mode. A
modular PhysicsList was used; HadronElasticPhysicsHP was enabled for
elastic neutron scattering and G4HadronPhysicsQGSP\_BIC\_HP for
inelastic hadronic processes. The high-precision neutron-transport
models used the G4NDL 4.7.1 nuclear-data library. Ion elastic/inelastic,
gamma-nuclear, electromagnetic, and decay processes were also included.

To test sensitivity to the internal target structure, two configurations
with the same total tungsten and water thicknesses were considered.
RADEX1 is a monolithic W 8.00 cm + H\textsubscript{2}O 4.00 cm geometry, whereas RADEX2
is the multilayer assembly {[}W 2.00 cm + H\textsubscript{2}O 0.28 cm{]}\ensuremath{\times}4 + H\textsubscript{2}O 2.88 cm
used in the RADEX neutron source. The total longitudinal thickness is
12.00 cm in both cases. All target layers had the same transverse
dimensions of 20 \ensuremath{\times} 20 cm\textsuperscript{2}. Tungsten (W) and water (H\textsubscript{2}O)
were defined using the standard Geant4 materials (G4\_W~and~G4\_WATER)
with their default compositions and densities. The global production
range cut was set to 0.1 mm using the command~/run/setCut 0.1 mm. The
primary proton beam was modeled as an ideal point beam with zero angular
divergence, directed along the longitudinal axis of the target. No
reflective or periodic boundary conditions were applied; particles
leaving the Geant4 world volume were removed from further tracking.

In the following, the term ``RADEX time-resolution function'' refers
specifically to the internal time response of the simulated
target-moderator assembly: the distribution of transport delays of
outgoing neutrons arising from their production, slowing down, and
transport within the W--H\textsubscript{2}O geometry considered here. This definition is
not identical to the full resolution function of a specific TOF
facility, which may additionally include the time structure of the
primary beam, the geometry and length of the flight path, angular
acceptance, sample dimensions, detector response, and electronic timing
uncertainty. These external contributions are not included in the
present Geant4 model of RADEX.

This comparison makes it possible to separate the effect of internal
material layering from that of the total material amount. The
simulations were performed for 10\textsuperscript{8} primary protons with an energy of 267
MeV. Neutrons with energies of 50--300 eV at all emission angles were
included in the analysis; 1 695 573 and 1 653 839 outgoing-neutron
records were obtained for RADEX1 and RADEX2, respectively. The use of
``all angles'' is deliberate: the aim is to determine the intrinsic time
response of the target-moderator assembly with maximum statistical
precision, without imposing the geometrical acceptance of a particular
experimental beam line. Angular selection is therefore not treated as
part of the RADEX function defined here. When the resulting function is
applied to a specific TOF facility, its angular acceptance (or the
corresponding weighting) must be introduced separately and checked for
possible changes in the time-distribution shape.

The resolution-function sample included neutrons produced in tungsten
that crossed the boundary of the final water layer for the first time in
the direction out of the target. For each track, the exit energy E\textsubscript{j} and
the full time delay T\textsubscript{j} were stored at this boundary.

Rare events with a large neutron-birth delay form a separate delayed
population. To construct the main time-resolution function, a prompt
sample was used with the neutron birth time measured from the start of
the primary proton event \(t_{birth} < 1\) \ensuremath{\mu}s. Inspection of the
original Geant4 records showed that the neutron birth time is measured
from the start of the same event initiated by the primary proton.

In the 50--300 eV range, RADEX1 contains 1 695 573 records, of which 14
are delayed and 1 695 559 are prompt; RADEX2 contains 1 653 839 records,
with 22 delayed and 1 653 817 prompt. The ordinary prompt population
ends at about 265 ns for RADEX1 and 175 ns for RADEX2, whereas the first
isolated delayed events appear only on the millisecond scale. The two
populations are separated by a time gap of about four orders of
magnitude. Because no events occur within this gap, any threshold placed
between the prompt and delayed populations yields the same main sample
and does not affect the determination of W2\_QUAD and G3. Each prompt
record \(\left( E_{j},T_{j} \right)\) is subsequently treated as an
individual observation in the Geant4 event sample.

To remove the leading energy dependence of the slowing-down time, the
dimensionless delay x\textsubscript{j}=T\textsubscript{j}/t\textsubscript{s}(E\textsubscript{j}), with t\textsubscript{s}(E)=0.5/\ensuremath{\sqrt{E}} \ensuremath{\mu}s and E in eV, is
used for each prompt event. After this scaling, the distribution shape
itself is compared rather than the trivial change of the overall time
scale with energy. In the statistical fit, E\textsubscript{j} is a known quantity
obtained from Geant4 and serves as a conditioning variable: the
distribution p(x\textbar E) is fitted, not the outgoing-neutron energy
spectrum dN/dE itself. For local analysis, events are grouped into the
intervals 50--100, 100--150, 150--200, 200--250, and 250--300 eV; for
the study of the rare far tail, the first interval is additionally
divided into 50--75 and 75--100 eV. Individual E\textsubscript{j} values of all prompt
events are used in the W2\_QUAD fit.

The main parameters of both geometries and the sizes of the
corresponding Geant4 event samples are summarized in Table 1.

\begin{longtable}{@{}
  >{\raggedright\arraybackslash}p{(\columnwidth - 4\tabcolsep) * \real{0.3334}}
  >{\raggedright\arraybackslash}p{(\columnwidth - 4\tabcolsep) * \real{0.3333}}
  >{\raggedright\arraybackslash}p{(\columnwidth - 4\tabcolsep) * \real{0.3333}}@{}}
\caption{Geant4 geometries used for the controlled comparison of RADEX1 and RADEX2.}\label{tab:1}\\

\toprule\noalign{}
\begin{minipage}[b]{\linewidth}\raggedright
Parameter
\end{minipage} & \begin{minipage}[b]{\linewidth}\raggedright
\textbf{RADEX1}
\end{minipage} & \begin{minipage}[b]{\linewidth}\raggedright
\textbf{RADEX2}
\end{minipage} \\
\midrule\noalign{}
\endhead
\bottomrule\noalign{}
\endlastfoot
Target structure & W 8.00 cm + H\textsubscript{2}O 4.00 cm & {[}W 2.00 cm + H\textsubscript{2}O 0.28
cm{]} \ensuremath{\times}4 + H\textsubscript{2}O 2.88 cm \\
Total W & 8.00 cm & 8.00 cm \\
Total H\textsubscript{2}O & 4.00 cm & 4.00 cm \\
Total longitudinal thickness & 12.00 cm & 12.00 cm \\
Primary beam & 267 MeV p & 267 MeV p \\
Analyzed range & 50--300 eV & 50--300 eV \\
Emission angles & all & all \\
Records, 50--300 eV: all / prompt & 1 695 573 / 1 695 559 & 1 653 839 /
1 653 817 \\
Main difference & monolithic geometry & multilayer W/H\textsubscript{2}O geometry \\
\end{longtable}

\section{From Geant4 events to an analytical resolution function}
The numerical Geant4 sample contains the required information about the
transport response in the form of individual events (E\textsubscript{j},x\textsubscript{j}), but a
compact normalized analytical function is needed for subsequent
convolution with the resonance shape. To avoid mixing the data with
their approximation, W1, W2, W3, W2\_QUAD, and W2\_QUAD+G3 are used
below only as names of model classes, whereas
p\textsubscript{M}(x\textbar E; \ensuremath{\theta}\textsubscript{M}) denotes the
normalized probability density of the selected model M. Geant4 does not
enter p\textsubscript{M} as an additional term; the simulation
information enters the statistical problem through the actual E\textsubscript{j} and x\textsubscript{j}
values of each event.

For a given model M and parameter set \ensuremath{\theta}\textsubscript{M}, the value
p\textsubscript{M}(x\textsubscript{j}\textbar E\textsubscript{j}; \ensuremath{\theta}\textsubscript{M}) is evaluated at
the coordinates of each individual Geant4 event. The numerical Geant4
observations provide D\textsubscript{G4}=\{(E\textsubscript{j},x\textsubscript{j})\}, whereas
p\textsubscript{M}(x\textbar E;\ensuremath{\theta}\textsubscript{M}) is a testable
analytical hypothesis for the distribution of those observations.
Maximum likelihood selects the values of \ensuremath{\theta}\textsubscript{M} for which
the observed set of delays is most probable. The corresponding negative
log-likelihood is

\[\mathrm{NLL}(\theta_M\mid D_{\mathrm{G4}})=-\sum_{j=1}^{N}\ln p_M(x_j\mid E_j;\theta_M).\]

In this sum, the index j runs over all stored Geant4 neutron records.
Each record contributes its delay x\textsubscript{j} and outgoing energy E\textsubscript{j} to the NLL;
histogramming does not enter the objective function. For the local
W1--W3 models, the parameters \ensuremath{\theta}\textsubscript{M} are common to all
events in the selected energy interval. For W2\_QUAD, the energy E\textsubscript{j} of
each event directly determines n\textsubscript{i}(E\textsubscript{j}), k\textsubscript{i}(E\textsubscript{j}), and \ensuremath{\epsilon}\textsubscript{2}(E\textsubscript{j}). Model
selection should account both for the quality of the event-level
description and for the penalty associated with additional parameters.

The event-level NLL was minimized using the bounded L-BFGS-B algorithm.
The Gamma shape and rate parameters were optimized in logarithmic form,
which guarantees~\(n_{i} + 1 > 0\)~and~\(k_{i} > 0\), while the mixture
weights were represented by logits and normalized by the softmax
transformation. Six optimization runs were performed for each W1, W2,
and W3 fit. The first run used moment-based initial estimates with
component means distributed over progressively longer time scales; the
remaining runs used reproducible random perturbations of these initial
parameters. The solution with the lowest finite NLL was retained. To
remove the arbitrary labeling of mixture components, the fitted
components were ordered by increasing mean
delay,~\(\left\langle x \right\rangle_{i} = \left( n_{i} + 1 \right)/k_{i}\).
The same bounded L-BFGS-B procedure and six restarts were used for the
constant-W2 and W2\_QUAD fits. W2\_QUAD was initialized by extending the
fitted constant-W2 solution with zero initial slope and curvature
coefficients. Broad finite parameter bounds were imposed only to
preserve the physical domains and prevent numerical overflow.

For model M, \ensuremath{\beta}\textsubscript{M} = dim(\ensuremath{\theta}\textsubscript{M}) denotes the
number of free parameters, and N is the number of Geant4 neutron records
in the fitted sample. The corresponding information criteria are

\[\mathrm{AIC}=2\beta_M+2\mathrm{NLL},\qquad \mathrm{BIC}=\beta_M\ln N+2\mathrm{NLL}.\]

AIC and BIC are used for comparative model selection among models
constructed from the same Geant4 sample. For a visual check of where the
analytical function departs from the transport distribution, the same
events are additionally grouped into histogram bins in x after fitting.
In principle, one primary proton can produce several outgoing-neutron
records. A clustering check for the RADEX2 prompt sample showed that 1
653 817 neutron records correspond to 1 607 247 distinct primary proton
events; the fraction of events with multiplicity \(> 1\) is 2.83\%.
Therefore, AIC and BIC are used below primarily as comparative criteria
for the distribution shape. An additional multiplicity check for the
rare far tail is given in Sec. 6. For mixture-component extensions,
where an added weight approaches zero at the boundary of the parameter
space, AIC and BIC are interpreted as comparative measures within the
specified family of candidate models rather than as a formal
significance test for the existence of the additional component.

Let n\textsubscript{i} be the observed number of Geant4 events in bin i with boundaries
{[}x\textsubscript{i}\textsubscript{1}, x\textsubscript{i}\textsubscript{2}{]}. The number of events expected from the model is
calculated from the same conditional density pM, taking the individual
energy of each event into account:

\[\mu_i(\theta_M)=\sum_{j=1}^{N}\int_{x_{i,1}}^{x_{i,2}}p_M(x\mid E_j;\theta_M)\,dx,\qquad r_i=\frac{n_i-\mu_i}{\sqrt{\mu_i}}.\]

Thus, the event-level NLL is the primary fit statistic and uses the
Geant4 information without intermediate histogramming. Histograms,
Pearson residuals, and \ensuremath{\chi}\textsuperscript{2}/Ndof are used only as shape diagnostics; in
the rare tail, \ensuremath{\chi}\textsuperscript{2} is not treated as an independent argument for the
statistical significance of G3. After the local structure of the
resolution function has been determined, its energy dependence is
described by a single smooth W2\_QUAD model. The complete set of
coefficients required to reproduce this model is given in Appendix A.

\section{Energy dependence of the main resolution-function shape}
\subsection{Time scaling and residual energy evolution}

Figure 1 shows how the shape of the dimensionless distribution
\(x = T/t_{s}(E)\) changes over the full 50--300 eV range. The Geant4 distributions in six energy intervals are compared directly, without analytical approximation. The scaling
\(t_{s}(E) = 0.5/\sqrt{E}\) \ensuremath{\mu}s brings the maxima and the main parts of
the distributions closer together, confirming the leading \(E^{- 1/2}\)
dependence of the time scale. At the same time, systematic energy
evolution remains in the far-tail region. This shows that a single
energy-independent shape is insufficient and motivates the construction
of a smooth energy-dependent W2\_QUAD model.

\begin{figure}[htbp]
\centering
\includegraphics[width=\linewidth]{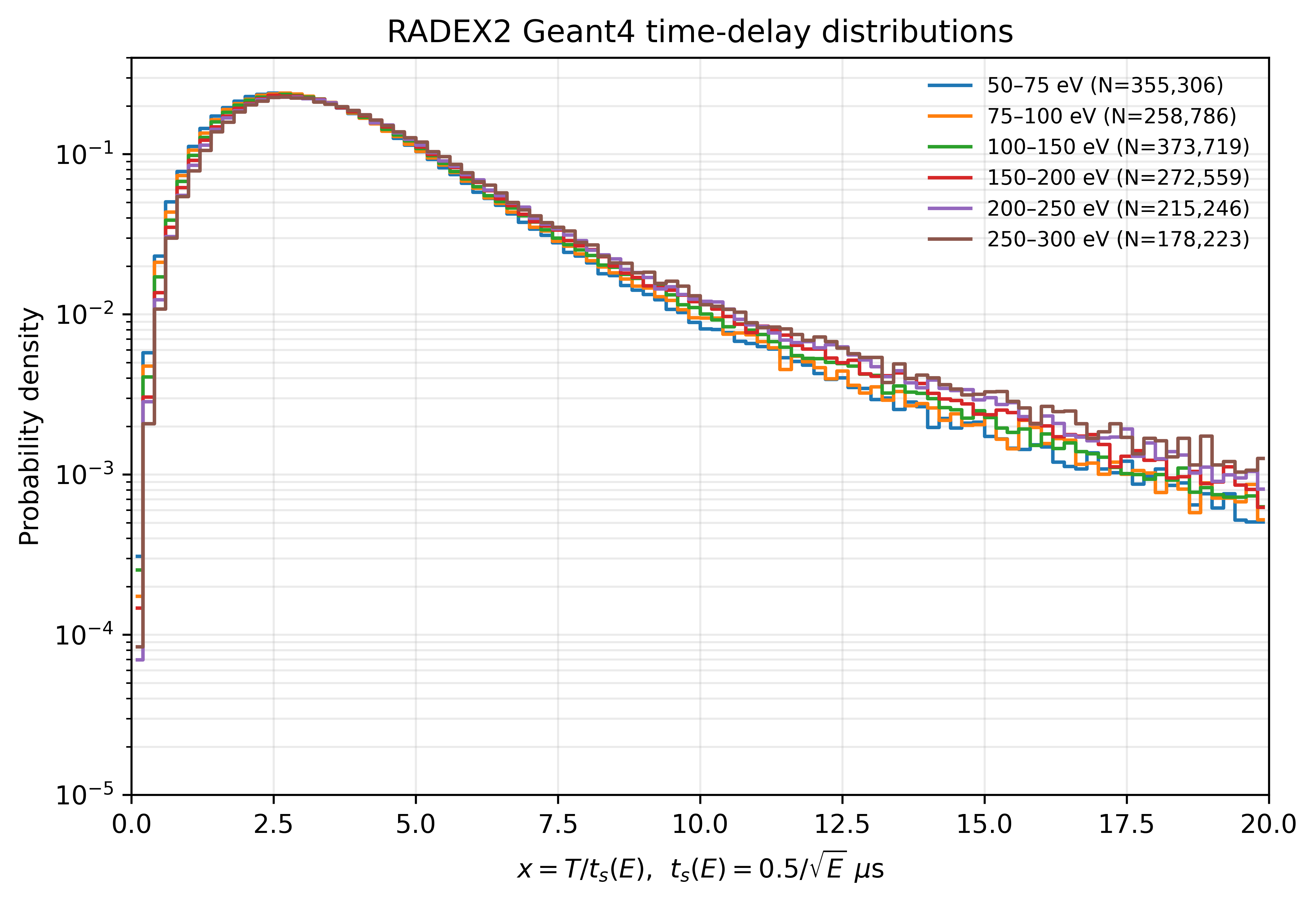}
\caption{Normalized distributions of the dimensionless variable \(x = T/t_{s}(E)\) for the multilayer RADEX2 geometry in six energy intervals. The main prompt sample with \(t_{birth} < 1\) \ensuremath{\mu}s is used and contains 1 653 817 neutron records. The scaling brings the main parts of the distributions closer together, whereas the logarithmic scale reveals the residual energy evolution of the far tail.}
\label{fig:1}
\end{figure}

\subsection{Construction of the smooth W2\_QUAD model}

After the main scale t\textsubscript{s}(E) has been removed, a weaker but systematic
change in the distribution shape remains. The model is described by the
two-component conditional density p\textsubscript{W2Q}(x\textbar E)
corresponding to W2\_QUAD:

\[p_{\mathrm{W2Q}}(x\mid E)=[1-\varepsilon_2(E)]g_1(x\mid E)+\varepsilon_2(E)g_2(x\mid E),\]

\[g_i(x\mid E)=\frac{k_i(E)[k_i(E)x]^{n_i(E)}e^{-k_i(E)x}}{\Gamma[n_i(E)+1]},\qquad \langle x\rangle_i(E)=\frac{n_i(E)+1}{k_i(E)}.\]

Independent W2 fits in separate energy intervals show that the
parameters vary smoothly with energy, but not strictly linearly. This
behavior is especially clear for the slow component: n\textsubscript{2}(E) and k\textsubscript{2}(E)
exhibit broad extrema, and \ensuremath{\epsilon}\textsubscript{2}(E) shows appreciable curvature as the
energy increases.

At a preliminary stage, a linear energy dependence of the parameters was
tested, but it did not reproduce the observed curvature, especially for
the slow-component parameters \(n_{2}(E)\), \(k_{2}(E)\) and the weight
\(\varepsilon_{2}(E)\), as seen directly in Fig. 2. A quadratic
dependence on the scaled energy u was therefore chosen for the
description:

\[u=\frac{E-E_{\mathrm{ref}}}{E_{\mathrm{scale}}},\qquad E_{\mathrm{ref}}=175\,\mathrm{eV},\quad E_{\mathrm{scale}}=125\,\mathrm{eV},\]

\[\ln[n_i(E)+1]=a_{i0}+a_{i1}u+a_{i2}u^2,\qquad \ln k_i(E)=b_{i0}+b_{i1}u+b_{i2}u^2,\]

\[\operatorname{logit}\varepsilon_2(E)=c_0+c_1u+c_2u^2.\]

The parameter transformations automatically preserve the physical
domains n\textsubscript{i} \textgreater{} \ensuremath{-}1, k\textsubscript{i} \textgreater{} 0, and 0 \textless{} \ensuremath{\epsilon}\textsubscript{2}
\textless{} 1. The analytical density pW2Q with this energy dependence
is referred to below as the W2\_QUAD model. The same two-component form
with parameters fixed within an individual energy interval corresponds
to the local W2 model. Unlike in earlier versions of the manuscript, the
complete set of 15 W2\_QUAD coefficients is not treated as an internal
technical detail: it is given in Appendix A together with the
definitions of u and t\textsubscript{s}(E) and the 50--300 eV range.

\begin{figure}[htbp]
\centering
\includegraphics[width=\linewidth]{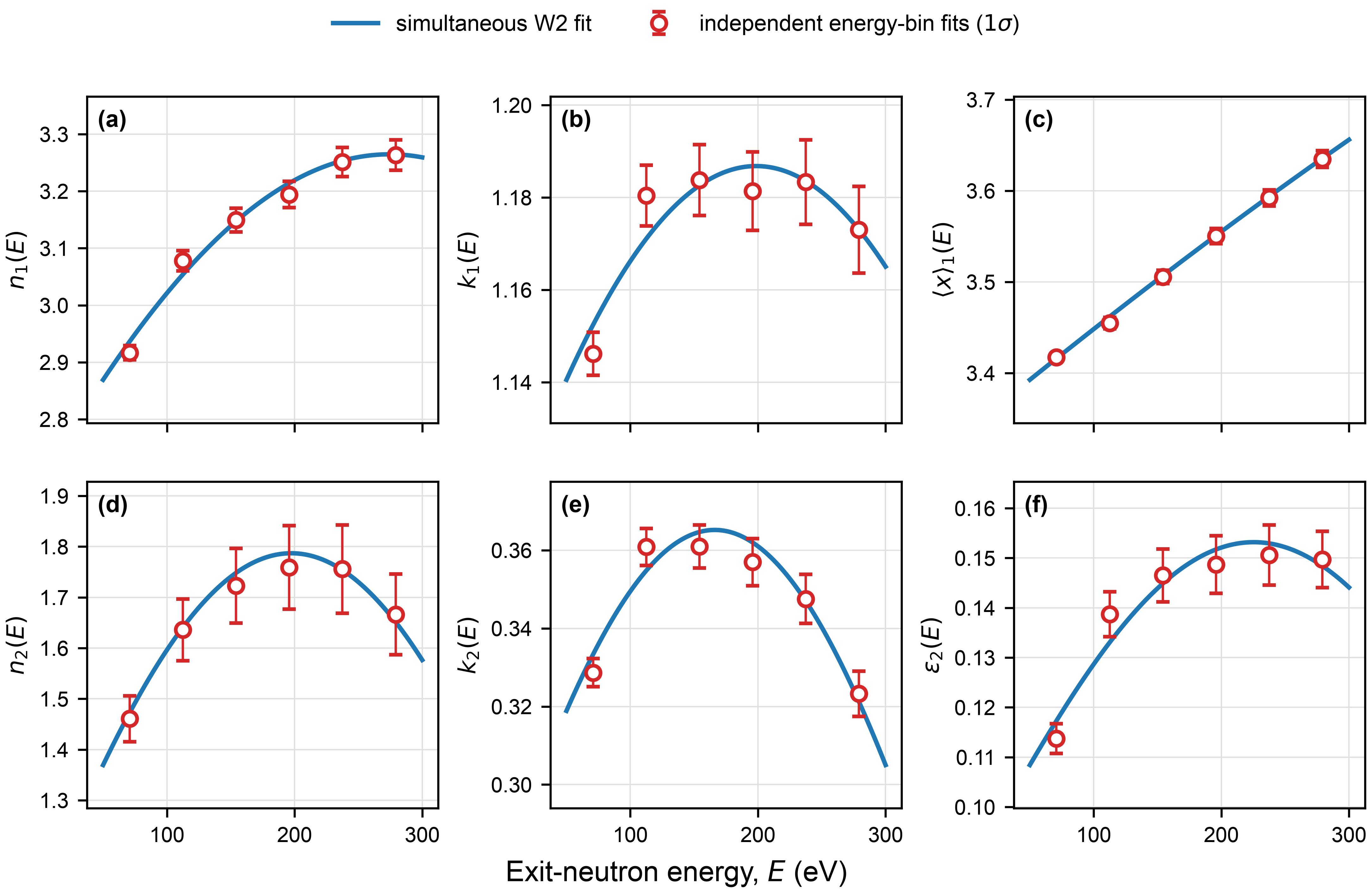}
\caption{Energy dependence of the parameters of the normalized two-component W2 model for multilayer RADEX2. Solid curves correspond to the joint event-level maximum-likelihood W2\_QUAD fit to 1 653 817 events of the main Geant4 sample over 50--300 eV; open circles show independent W2 fits in six energy intervals, with 1\ensuremath{\sigma} error bars. Relative to an energy-independent W2 model, the quadratic parametrization improves the description by \ensuremath{\Delta}NLL=5180.988, \ensuremath{\Delta}AIC=10341.977, and \ensuremath{\Delta}BIC=10218.791.}
\label{fig:2}
\end{figure}

The open circles in Fig. 2 represent independent constant-W2
maximum-likelihood fits performed separately in the six energy
intervals. For each fit, the covariance matrix was obtained by inverting
the finite-difference Hessian of the event-level NLL evaluated at the
optimum. The reported~\(1\sigma\)~uncertainties of the physical and
derived parameters were calculated by first-order covariance propagation
(the delta method). A pseudoinverse was used when the Hessian was
numerically rank deficient.

A direct illustration of this construction is shown for the multilayer
RADEX2 geometry. The solid curves in Fig. 2 are obtained from a single
joint event-level maximum-likelihood fit of the main sample over the
full 50--300 eV range, whereas the points correspond to independent W2
fits in the six energy intervals and were not used as input data for
constructing the curves. The constant two-component W2 model already
describes the main asymmetric shape, but allowing its parameters to vary
smoothly with energy substantially improves the description.

For the multilayer RADEX2 target, the NLL, AIC, and BIC values for the
constant W2 model and W2\_QUAD are listed in Table 2.

Hereafter \ensuremath{\Delta}NLL, \ensuremath{\Delta}AIC, and \ensuremath{\Delta}BIC are defined as the value of the criterion for the simpler model minus its value for the extended model. Accordingly, a positive \ensuremath{\Delta} denotes an improvement in the description.  For RADEX2, the transition to W2\_QUAD reduces the NLL by 5180.988; despite the increase in the number of parameters from 5 to 15, the BIC decreases by 10218.791. Thus, describing the energy evolution of the main time-response shape requires the energy-dependent W2\_QUAD form.

\begin{longtable}{@{}
  >{\raggedright\arraybackslash}p{(\columnwidth - 8\tabcolsep) * \real{0.2000}}
  >{\raggedright\arraybackslash}p{(\columnwidth - 8\tabcolsep) * \real{0.2000}}
  >{\raggedright\arraybackslash}p{(\columnwidth - 8\tabcolsep) * \real{0.2000}}
  >{\raggedright\arraybackslash}p{(\columnwidth - 8\tabcolsep) * \real{0.2000}}
  >{\raggedright\arraybackslash}p{(\columnwidth - 8\tabcolsep) * \real{0.2000}}@{}}
\caption{Comparison of the constant W2 model and the energy-dependent W2\_QUAD model for the main RADEX2 sample. AIC/BIC are calculated with N equal to the number of neutron records and are used as internal comparative criteria.}\label{tab:2}\\

\toprule\noalign{}
\begin{minipage}[b]{\linewidth}\raggedright
Model
\end{minipage} & \begin{minipage}[b]{\linewidth}\raggedright
\textbf{\ensuremath{\beta}\textsubscript{M}}
\end{minipage} & \begin{minipage}[b]{\linewidth}\raggedright
\textbf{NLL}
\end{minipage} & \begin{minipage}[b]{\linewidth}\raggedright
\textbf{AIC}
\end{minipage} & \begin{minipage}[b]{\linewidth}\raggedright
\textbf{BIC}
\end{minipage} \\
\midrule\noalign{}
\endhead
\bottomrule\noalign{}
\endlastfoot
W2 constant & 5 & 9 806 806.317 & 19 613 622.634 & 19 613 684.227 \\
W2\_QUAD & 15 & 9 801 625.329 & 19 603 280.658 & 19 603 465.436 \\
Improvement & +10 & 5 180.988 & 10 341.977 & 10 218.791 \\
\end{longtable}

\section{How many time scales are required to describe the spectrum?}
Having established the energy-dependent main shape, we examine how many
Gamma components are needed to describe a local time spectrum. The
transition W1\ensuremath{\to}W2\ensuremath{\to}W3 is illustrated in the
200--250 eV interval; the additional component is tested separately in
Sec. 6 while holding the W2\_QUAD main shape fixed.

\begin{figure}[htbp]
\centering
\includegraphics[width=\linewidth]{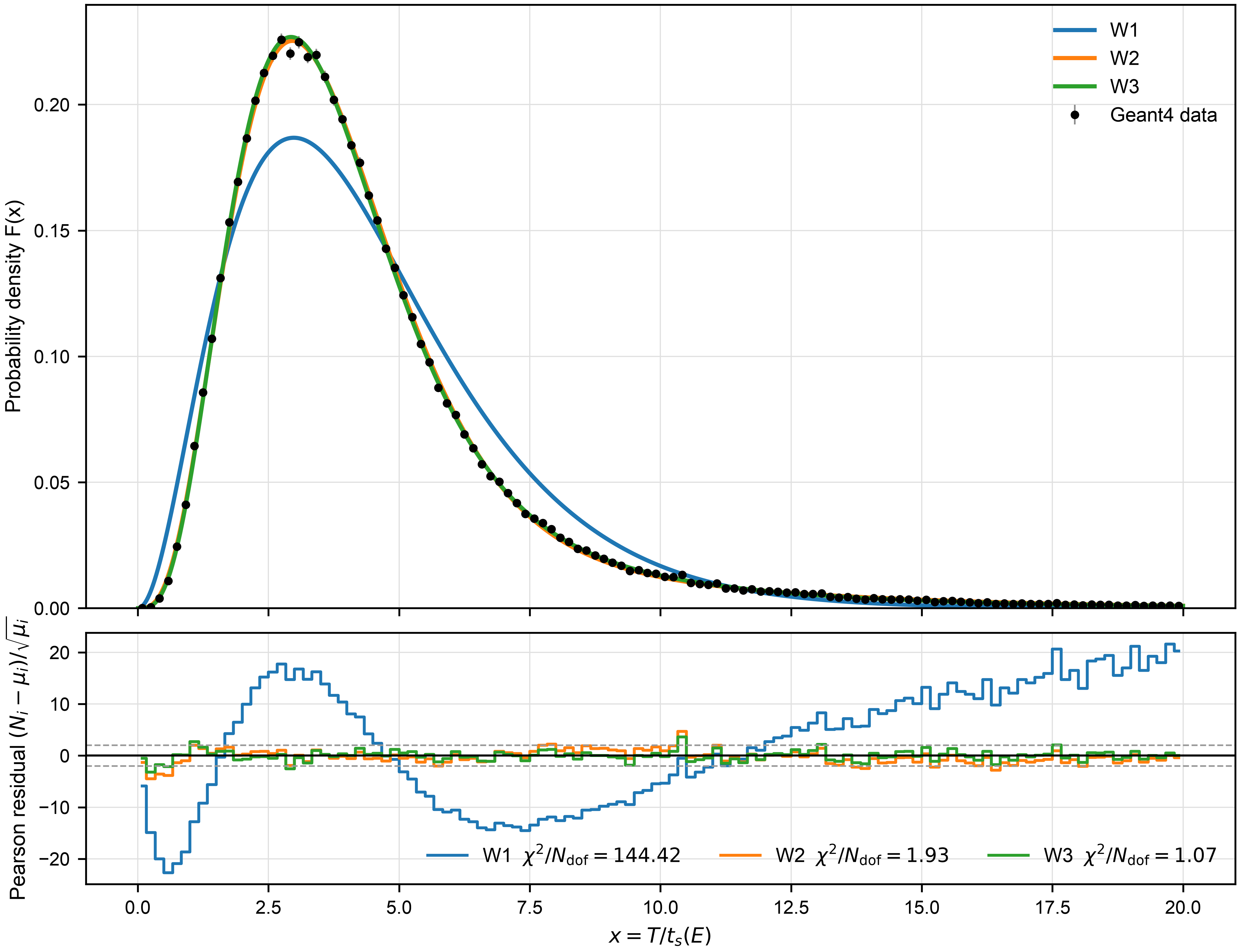}
\caption{Comparison of the one-, two-, and three-component Gamma models W1, W2, and W3 with the main Geant4 sample for RADEX2 in the 200--250 eV interval. The upper panel shows the probability density and the lower panel the Pearson residuals. The resulting values are \ensuremath{\chi}\textsuperscript{2}/Ndof=144.421 for W1, 1.929 for W2, and 1.066 for W3.}
\label{fig:3}
\end{figure}

For the 200--250 eV example (Fig. 3), W1 clearly fails to reproduce the
shape of the time distribution: \ensuremath{\chi}\textsuperscript{2}/Ndof=144.421. W2 reduces this value
to 1.929 and W3 to 1.066; \ensuremath{\Delta}BIC(W2\ensuremath{-}W3)=134.87.

Thus, in the 200--250 eV interval, W3 provides a substantially better
description of the displayed distribution and motivates the more
stringent far-tail test presented in Sec. 6.

The parameters of the first two components are noticeably correlated and
may vary between energy intervals. The physically relevant feature is
therefore not the ``jump'' of an individual n\textsubscript{1}, k\textsubscript{1}, n\textsubscript{2}, or k\textsubscript{2} parameter,
but the stable property of the complete shape: the presence of an
additional slow time scale, confirmed by the analysis of the full time
variable in Fig. 3.

\FloatBarrier
\section{Long-time tail: the W2\_QUAD+G3 extension}
Local W3 fits show that a third time scale is required in the tail
region. However, in a local W3 model all three Gamma components are
reoptimized simultaneously, so the improvement by itself does not
distinguish a genuinely new tail component from a correlated
readjustment of the first two components. A more stringent test is
therefore used: the already determined energy-dependent main shape
W2\_QUAD, i.e. the density p\textsubscript{W2Q}(x\textbar E), is first
fixed, and only one additional Gamma component, G3, is then added. Its
shape is described by two parameters n\textsubscript{3} and k\textsubscript{3} common to all energy
intervals, while only the small weight \ensuremath{\epsilon}\textsubscript{3}(E) varies between intervals.
The normalized density of the extended W2\_QUAD+G3 model is denoted by \(p_{\mathrm{ext}}(x\mid E)\):

\[p_{\mathrm{ext}}(x\mid E)=[1-\varepsilon_3(E)]p_{\mathrm{W2Q}}(x\mid E)+\varepsilon_3(E)g_3(x;n_3,k_3).\]

This test is intentionally more stringent than an independent local W3
fit: all 15 W2\_QUAD coefficients remain fixed and cannot readjust to
absorb the residual.

In the joint event-level maximum-likelihood fit, only two common G3
shape parameters (\(n_{3},k_{3}\)) and six interval-dependent
weights~\(\varepsilon_{3}\)~were optimized, giving a total of eight
additional parameters. Four reproducible L-BFGS-B runs were performed,
and the solution with the lowest finite NLL was retained. To exclude a
degenerate solution and prevent interchange of G3 with the components of
the fixed W2\_QUAD model, the mean G3 delay was required to exceed the
maximum mean delay of the slow W2\_QUAD component by at least 0.25 in
the variable~\(x\). This constraint only separates the components and
does not predetermine the position of G3.

For RADEX2, the joint fit gives \ensuremath{\langle}x\ensuremath{\rangle}\textsubscript{3}=30.7, well beyond the main W2\_QUAD
distribution. The event-level NLL is evaluated over the full prompt
sample using the individual E\textsubscript{j} and x\textsubscript{j} of each event. The results for the
six energy intervals are listed in Table 3.

\begin{longtable}{@{}
  >{\raggedright\arraybackslash}p{(\columnwidth - 10\tabcolsep) * \real{0.1667}}
  >{\raggedright\arraybackslash}p{(\columnwidth - 10\tabcolsep) * \real{0.1667}}
  >{\raggedright\arraybackslash}p{(\columnwidth - 10\tabcolsep) * \real{0.1667}}
  >{\raggedright\arraybackslash}p{(\columnwidth - 10\tabcolsep) * \real{0.1667}}
  >{\raggedright\arraybackslash}p{(\columnwidth - 10\tabcolsep) * \real{0.1667}}
  >{\raggedright\arraybackslash}p{(\columnwidth - 10\tabcolsep) * \real{0.1667}}@{}}
\caption{Joint test of the additional Gamma component G3 for RADEX2: comparison of the fixed W2\_QUAD model with the extended W2\_QUAD+G3 model in six energy intervals of the prompt sample. The event-level NLL is evaluated over the full sample; \ensuremath{\chi}\textsuperscript{2}/Ndof is used as a histogram-based diagnostic with 160 bins over 0\ensuremath{\leq}x\ensuremath{\leq}40.}\\

\toprule\noalign{}
\begin{minipage}[b]{\linewidth}\raggedright
Interval, eV
\end{minipage} & \begin{minipage}[b]{\linewidth}\raggedright
\textbf{N}
\end{minipage} & \begin{minipage}[b]{\linewidth}\raggedright
\textbf{\ensuremath{\epsilon}\textsubscript{3}, \%}
\end{minipage} & \begin{minipage}[b]{\linewidth}\raggedright
\textbf{\ensuremath{\Delta}NLL}
\end{minipage} & \begin{minipage}[b]{\linewidth}\raggedright
\ensuremath{\chi}\textsuperscript{2}/Ndof W2\_QUAD
\end{minipage} & \begin{minipage}[b]{\linewidth}\raggedright
\ensuremath{\chi}\textsuperscript{2}/Ndof W2\_QUAD+G3
\end{minipage} \\
\midrule\noalign{}
\endhead
\bottomrule\noalign{}
\endlastfoot
50--75 & 355 304 & 0.05253 & 89.831 & 3.905 & 2.399 \\
75--100 & 258 780 & 0.03932 & 42.492 & 3.054 & 1.866 \\
100--150 & 373 712 & 0.03611 & 53.300 & 3.449 & 2.216 \\
150--200 & 272 554 & 0.03439 & 33.179 & 2.655 & 1.825 \\
200--250 & 215 245 & 0.04764 & 31.072 & 2.582 & 1.751 \\
250--300 & 178 222 & 0.04951 & 19.637 & 2.064 & 1.625 \\
50--300 & 1 653 817 & 0.034--0.053 & 269.511 & 2.952 & 1.951 \\
\end{longtable}

The interval-specific \ensuremath{\chi}\textsuperscript{2}/Ndof values are conditional on the common n\textsubscript{3}
and k\textsubscript{3} parameters: for each individual interval, only one independently
fitted parameter, \ensuremath{\epsilon}\textsubscript{3}, is subtracted in calculating Ndof. The final row
accounts for all eight additional parameters of the W2\_QUAD+G3 model.

n\textsubscript{3} = 20.3076, k\textsubscript{3} = 0.694841, \ensuremath{\langle}x\ensuremath{\rangle}\textsubscript{3} = 30.7

\ensuremath{\langle}\ensuremath{\Delta}t\ensuremath{\rangle}\textsubscript{3}(E) = \ensuremath{\langle}x\ensuremath{\rangle}\textsubscript{3} t\textsubscript{s}(E) = 15.3 / \ensuremath{\sqrt{E}} \ensuremath{\mu}s.

The resulting common scale corresponds to a mean delay of about 1.08 \ensuremath{\mu}s
at 200 eV. The integrated G3 weight remains very small---0.034--0.053\%
in the different energy intervals. This number is the weight of a
normalized mixture component and does not imply that all events
associated with G3 lie only at x\textgreater30. The contribution of G3
becomes especially visible in the far-tail region where the main
W2\_QUAD density is already exponentially suppressed.

The possible role of primary-event multiplicity in generating the far
tail was also examined. A check using the primary proton event
identifier showed that rare events in the long-time tail do not result
from multiple neutron records produced by the same primary proton. In
total, 88 neutron records were found in the region \(200 \leq E < 250\)
eV, \(x > 30\), corresponding to 88 distinct primary proton events.
Therefore, the observed tail component that motivates the introduction
of the additional Gamma component G3 cannot be explained by simple
repeated counting of several neutron records from one primary proton
event. This result shows that the far tail is a property of the transport time
distribution rather than an artifact of counting multiple neutrons from
the same primary proton event.

\begin{figure}[htbp]
\centering
\includegraphics[width=\linewidth]{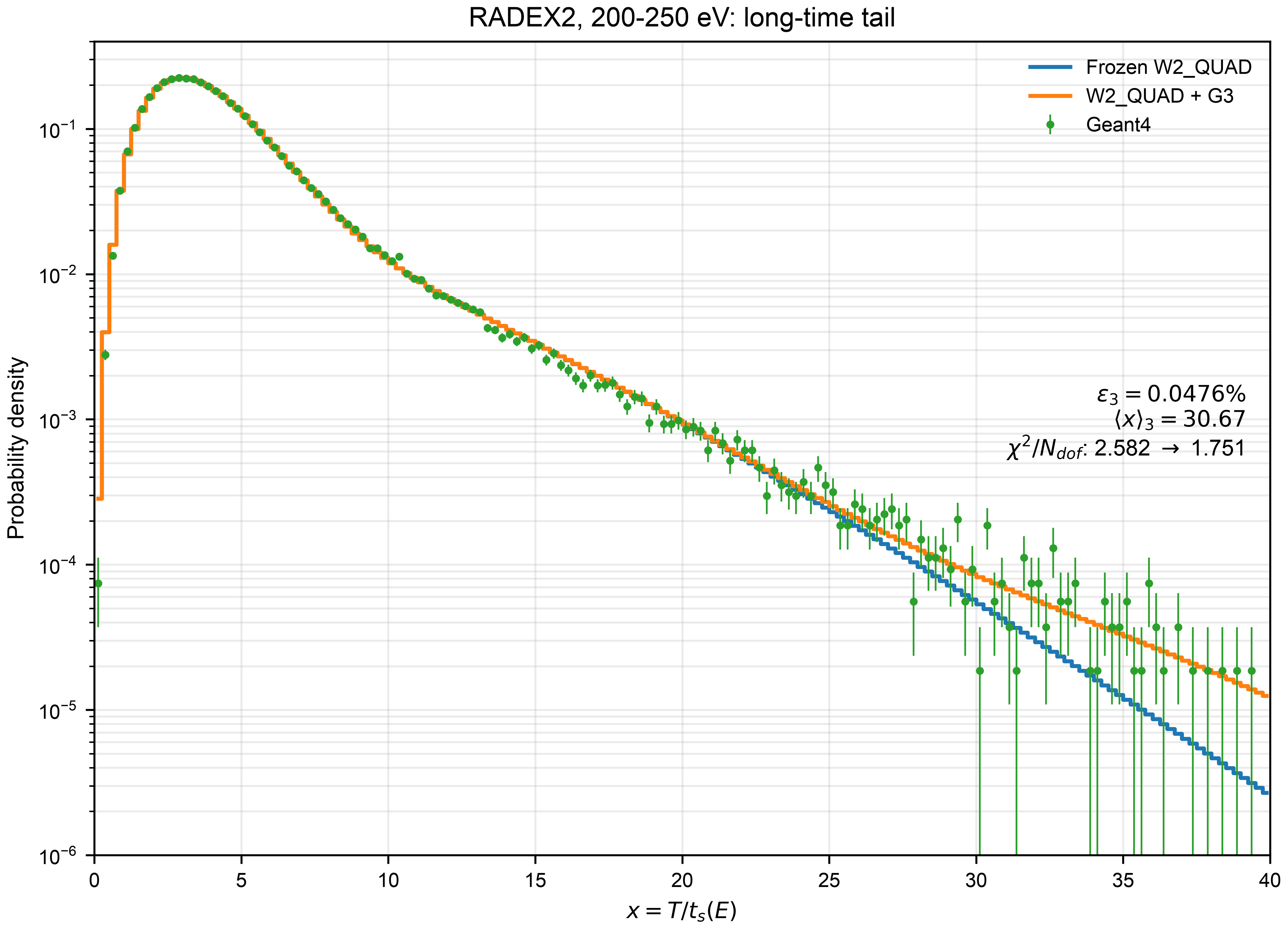}
\caption{Long-time tail of RADEX2 in the 200--250 eV interval for the prompt sample. The fixed W2\_QUAD model describes the main part of the distribution well but systematically underestimates the large-x region. Adding a common-shape G3 component with interval weight \ensuremath{\epsilon}\textsubscript{3}=0.04764\% and \ensuremath{\langle}x\ensuremath{\rangle}\textsubscript{3}=30.67 improves \ensuremath{\chi}\textsuperscript{2}/Ndof from 2.582 to 1.751. The plot shows x\ensuremath{\leq}40; the event-level maximum-likelihood fit is not restricted to this range.}
\label{fig:4}
\end{figure}

The joint fit of all six intervals gives \ensuremath{\Delta}NLL=269.511. After accounting
for eight additional parameters, \ensuremath{\Delta}AIC=523.021 and \ensuremath{\Delta}BIC=424.472 are
obtained in favor of W2\_QUAD+G3. These values refer to the prompt
sample of 1 653 817 neutron records; the final parameters and
information criteria refer exclusively to the analysis of the main
sample.

For the principal RADEX2 geometry, the W2\_QUAD model of the main
time-distribution shape was determined above by a joint event-level fit
over 50--300 eV. After its 15 coefficients had been fixed, the G3
component was added with two common shape parameters and a weight
determined independently in each of the six energy intervals.

To test the sensitivity of the far tail to the internal target
structure, the same procedure was applied independently to RADEX1. Its
separate 15-parameter W2\_QUAD model was fitted and fixed before the G3
test in the same six energy intervals. The RADEX2 coefficients were
therefore not transferred to RADEX1. RADEX1 is included solely as a
control geometry for comparing the calculated time distributions; its
W2\_QUAD coefficients are not reported because the principal object of
this study is the multilayer RADEX2 target.

The comparison suggests that the far tail of the time distribution is
sensitive to the internal target geometry. For the same total W and H\textsubscript{2}O
thicknesses, the point estimates of the G3 weight are higher for the
multilayer RADEX2 target than for RADEX1 in all six energy intervals.
This consistent difference suggests a more pronounced far tail for
RADEX2.

The estimated number of events corresponding to the G3 weight is
approximately 708 for RADEX2 and 172 for RADEX1. In both geometries, the
fitted G3 component corresponds to a distinct long-delay scale,
with~\(\left\langle x \right\rangle_{3} \approx 30.7\)~for RADEX2
and~\(\approx 38.3\)~for RADEX1.

The numerical results of this comparison are summarized in Table 4,
which lists the G3 parameters, fit-quality measures, and its weight
\(\varepsilon_{3}\) in all energy intervals. Since uncertainties of the
small interval-dependent G3 weights were not evaluated, the
RADEX2--RADEX1 comparison is based on the consistency of the point
estimates across the six energy intervals and is not interpreted as a
formal test of statistical significance.

\begin{longtable}{@{}
  >{\raggedright\arraybackslash}p{(\columnwidth - 4\tabcolsep) * \real{0.3334}}
  >{\raggedright\arraybackslash}p{(\columnwidth - 4\tabcolsep) * \real{0.3333}}
  >{\raggedright\arraybackslash}p{(\columnwidth - 4\tabcolsep) * \real{0.3333}}@{}}
\caption{Final characteristics of the additional Gamma component G3 for RADEX2 and RADEX1. The \ensuremath{\epsilon}\textsubscript{3} values are point estimates.}\\

\toprule\noalign{}
\begin{minipage}[b]{\linewidth}\raggedright
Quantity
\end{minipage} & \begin{minipage}[b]{\linewidth}\raggedright
\textbf{RADEX2}
\end{minipage} & \begin{minipage}[b]{\linewidth}\raggedright
\textbf{RADEX1}
\end{minipage} \\
\midrule\noalign{}
\endhead
\bottomrule\noalign{}
\endlastfoot
Number of records in the main sample & 1 653 817 & 1 695 559 \\
n\textsubscript{3} & 20.3076 & 12.5643 \\
k\textsubscript{3} & 0.694841 & 0.353932 \\
\ensuremath{\langle}x\ensuremath{\rangle}\textsubscript{3}=(n\textsubscript{3}+1)/k\textsubscript{3} & 30.7 & 38.3 \\
Range of \ensuremath{\epsilon}\textsubscript{3}(E), \%, point estimates & 0.03439--0.05253 &
0.00537--0.01482 \\
Estimated number of G3-component events, \ensuremath{\Sigma} N\_z \ensuremath{\epsilon}\textsubscript{3},z & \ensuremath{\approx}708 & \ensuremath{\approx}172 \\
\ensuremath{\Delta}NLL & 269.511 & 142.409 \\
\ensuremath{\Delta}AIC & 523.021 & 268.818 \\
\ensuremath{\Delta}BIC & 424.472 & 170.070 \\
\ensuremath{\chi}\textsuperscript{2}/Ndof, fixed W2\_QUAD (x\ensuremath{\leq}40) & 2.952 & 1.911 \\
\ensuremath{\chi}\textsuperscript{2}/Ndof, W2\_QUAD+G3 (x\ensuremath{\leq}40) & 1.951 & 1.621 \\
\ensuremath{\epsilon}\textsubscript{3}, 50--75 eV, \% & 0.05253 & 0.01266 \\
\ensuremath{\epsilon}\textsubscript{3}, 75--100 eV, \% & 0.03932 & 0.00648 \\
\ensuremath{\epsilon}\textsubscript{3}, 100--150 eV, \% & 0.03611 & 0.01104 \\
\ensuremath{\epsilon}\textsubscript{3}, 150--200 eV, \% & 0.03439 & 0.00537 \\
\ensuremath{\epsilon}\textsubscript{3}, 200--250 eV, \% & 0.04764 & 0.01098 \\
\ensuremath{\epsilon}\textsubscript{3}, 250--300 eV, \% & 0.04951 & 0.01482 \\
\end{longtable}

\section{Comparison with the calculated n\_TOF resolution function}
After establishing a three-component structure for RADEX2, a natural
question arises: can the same analytical scheme be applied to the
resolution function of a source with a different design? For this
comparison, the SAMMY Tr108 calculated function {[}4,5{]} is used for
the Pb+H\textsubscript{2}O system at a flight path of L=183.9 m. The analytical approach
is applicable to n\_TOF as well, but the resolution-function parameters
for RADEX and n\_TOF must be determined independently. For n\_TOF, a
separate definition of the zero point and the dimensionless variable x
is used, as described below.

For each nominal energy E\textsubscript{0}=150, 200, 250, and 300 eV, the original Tr108
function is provided by SAMMY as a calculated density on an energy grid
E. This grid is first converted to time of flight using
t(E)=72.29825\ensuremath{\cdot}L/\ensuremath{\sqrt{E}} \ensuremath{\mu}s at L=183.9 m. The delay zero point is set by the
onset energy Eonset of the corresponding Dirac response: 150.08, 200.12,
250.15, and 300.18 eV. The digits after the decimal point indicate the
coordinate of the selected point on the original SAMMY energy grid and
do not represent the physical accuracy with which the boundary is
determined.

The variable x={[}t(E)\ensuremath{-}t(Eonset){]}/t\textsubscript{s}(E), with t\textsubscript{s}(E)=0.5/\ensuremath{\sqrt{E}} \ensuremath{\mu}s, is then
introduced. Here \(E\) denotes the energy of the current Tr108 grid
point rather than the nominal energy \(E_{0}\). Therefore, the scale
\(t_{s}(E)\) is evaluated separately at each point, as in the processing
of the Geant4 distributions. This preserves the same definition of the
dimensionless variable \(x\) for RADEX and n\_TOF. For each nominal
energy \(E_{0}\), the value \(E_{\text{onset}}\) was selected
graphically from the corresponding SAMMY Dirac response as its
high-energy boundary, i.e. the maximum energy at which the calculated
response is still nonzero. This boundary corresponds to the earliest
arrival time. Thus, x=0 corresponds to the leading edge of the
calculated Tr108 response, rather than to the distribution maximum or to
a fitted time shift.

When converting the SAMMY density in energy to a density in x, the
Jacobian \textbar dE/dx\textbar{} is included. The resulting function is
interpolated onto a uniform grid 0\ensuremath{\leq}x\ensuremath{\leq}60 (6001 points) and normalized to
unit area; denote it by p(x). Each W2 or W3 model on the same grid is
likewise normalized to unit area and defines q(x). The quantitative
measure of shape difference is the Kullback--Leibler divergence
KL(p\textbar\textbar q)=\ensuremath{\int}p(x) ln{[}p(x)/q(x){]} dx. Numerically, the
integral is evaluated on the same x grid using the trapezoidal rule. In
regions where the density is effectively zero, a small positive floor
(10\textsuperscript{-300}) is used in the logarithm to prevent log(0);
it does not modify the normalized densities \(p(x)\) and \(q(x)\) and is
used only for numerical stabilization. KL\ensuremath{\geq}0, and KL=0 corresponds to
identical normalized shapes.

\begin{figure}[htbp]
\centering
\includegraphics[width=\linewidth]{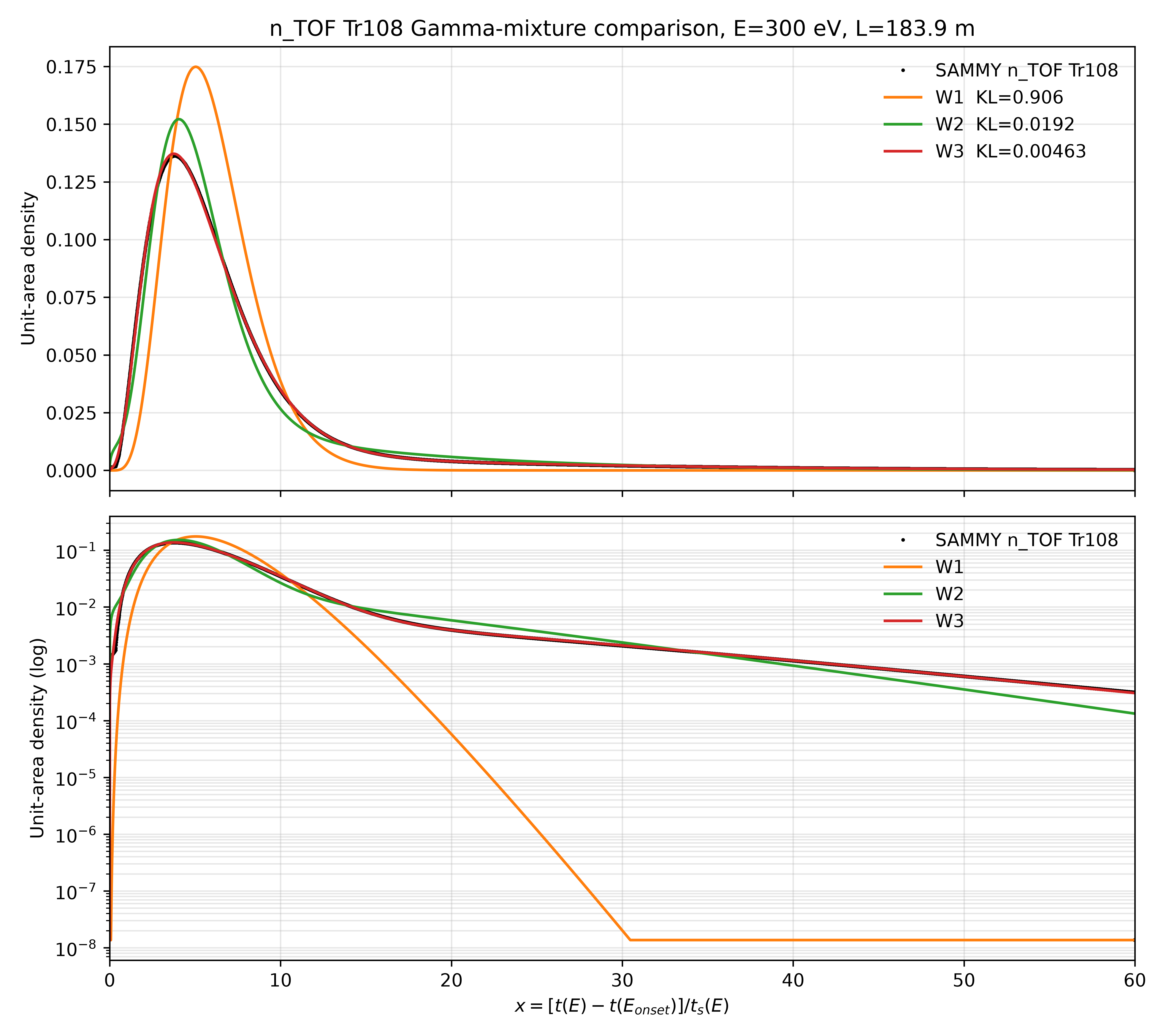}
\caption{Normalized calculated n\_TOF Tr108 function at E=300 eV and L=183.9 m compared with W1, W2, and W3. The linear scale shows the main shape and the logarithmic scale the far tail. At 300 eV, KL(W2)=0.01915 and KL(W3)=0.004628.}
\label{fig:5}
\end{figure}

At each energy, the calculated n\_TOF function was fitted independently
with W2 and W3 using the same variable \(x\) and the same normalization.
The Gamma-component parameters were obtained by minimizing
\(KL(p \parallel q)\) between the calculated distribution \(p(x)\) and
the model \(q(x)\).

At all four energies, adding the third Gamma component reduces KL and
improves the description of the tail region. Figure 5 shows the
calculated function at 300 eV: on the linear scale W2 already reproduces
the main maximum well, whereas the logarithmic scale reveals a
systematic discrepancy in the tail; W3 describes this region more
accurately.

\begin{longtable}{@{}
  >{\raggedright\arraybackslash}p{(\columnwidth - 6\tabcolsep) * \real{0.2500}}
  >{\raggedright\arraybackslash}p{(\columnwidth - 6\tabcolsep) * \real{0.2500}}
  >{\raggedright\arraybackslash}p{(\columnwidth - 6\tabcolsep) * \real{0.2500}}
  >{\raggedright\arraybackslash}p{(\columnwidth - 6\tabcolsep) * \real{0.2500}}@{}}
\caption{Comparison of two- and three-component descriptions of the calculated n\_TOF Tr108 function.}\label{tab:5}\\

\toprule\noalign{}
\begin{minipage}[b]{\linewidth}\raggedright
E, eV
\end{minipage} & \begin{minipage}[b]{\linewidth}\raggedright
\textbf{KL(W2)}
\end{minipage} & \begin{minipage}[b]{\linewidth}\raggedright
\textbf{KL(W3)}
\end{minipage} & \begin{minipage}[b]{\linewidth}\raggedright
\textbf{KL(W2)/KL(W3)}
\end{minipage} \\
\midrule\noalign{}
\endhead
\bottomrule\noalign{}
\endlastfoot
150 & 0.01179 & 0.001623 & 7.27 \\
200 & 0.01010 & 0.003297 & 3.06 \\
250 & 0.01265 & 0.003516 & 3.60 \\
300 & 0.01915 & 0.004628 & 4.14 \\
\end{longtable}

For n\_TOF, the main measure of shape quality is KL. Because Tr108 is a
calculated function rather than an independent event sample with a
defined N, a conditional BIC is not used here. Independent fits showed
that W3 describes the calculated n\_TOF function well at each energy.
The possibility of using one common W3 shape simultaneously at 150, 200,
250, and 300 eV was also tested. In this test, the shapes of the three
Gamma components were energy independent and only their relative weights
varied. The joint fit identified a common slow component with mean
dimensionless delay \(\left\langle x \right\rangle_{3} \approx 20.3\).
Its weight increases from 11.49\% at 150 eV to 15.42\% at 300 eV. This
supports the presence of a common slow time scale in the n\_TOF
function.

The onset energies were determined visually from the first appreciable
nonzero values of the resolution function and are therefore quoted only
to the precision justified by the source data. A small shift in the
onset position primarily changes the origin of the transformed variable
(x), whereas the observed long-delay tail extends over a substantially
broader range. Such a shift therefore does not alter the qualitative
conclusion that the three-component representation provides the best
description of the n\_TOF distributions. No quantitative sensitivity
analysis of the fitted parameters to the choice of onset energy was
performed.

This result is regarded only as evidence for the presence of a slow
component, rather than as a definitive model of the energy dependence of
the n\_TOF resolution function. The parameters of this component differ
from those obtained for RADEX2; therefore, the parameters cannot be
transferred between the two facilities.

\section{Discussion}
The results provide a consistent overall picture of the RADEX time
response. The main delay scale is determined by the slowing-down process
and approximately follows an E\textsuperscript{-}\textsuperscript{1}\ensuremath{/}\textsuperscript{2} dependence. The main part of the time
distribution is described by the W2\_QUAD model, whose parameters vary
smoothly with neutron energy. At large delays, however, this model
underestimates the number of events, so an additional slow G3 component
is required to describe the far tail.

Geant4 directly provides the time distribution of neutrons emerging from
the target. The W1, W2, W3, W2\_QUAD, and W2\_QUAD+G3 models are used
only as compact analytical descriptions of that distribution. The first
two Gamma components overlap strongly and can partially compensate each
other. Therefore, the values \(n_{1}\), \(k_{1}\), \(n_{2}\) and
\(k_{2}\) may vary appreciably while the resulting distribution shape
remains almost unchanged. These individual parameters should not be
assigned a direct physical meaning. The robust physical result is that
the distribution is asymmetric, its shape varies with energy, and an
additional slow G3 component is required to describe the far tail.

The comparison with RADEX1 suggests that the far-tail component is
determined not only by the total W and H\textsubscript{2}O thicknesses but may also
depend on their spatial arrangement.

Comparison with n\_TOF Tr108 supports the generality of the approach but
does not justify reconstructing real n\_TOF data using RADEX parameters.
The n\_TOF Tr108 function used here is a calculated realization of the
resolution function for a different Pb+H\textsubscript{2}O system. Reliable analysis of
experimental data requires the full time-response function of the
particular facility, including the source, target, moderator, flight
path, and detector.

A more general methodological conclusion follows. If the resolution
function is incomplete or inaccurate, the mismatch between the true and
assumed time response can be absorbed into the fitted resonance
parameters. Such an error is systematic. Determination, validation, and
documentation of the time-resolution function should therefore be
treated as an independent part of the experiment, rather than as an
auxiliary technical procedure performed only during subsequent resonance
reconstruction.

\section{Conclusions}
Geant4 was used to calculate the time distribution from the arrival of
the proton beam at the target to the neutron exit from it. This
distribution defines the time-resolution function of the RADEX source.
Neutron slowing down and transport in the target constitute a complex
stochastic process with several characteristic time scales. The
calculated time distribution is therefore approximated by a mixture of
Gamma functions describing the main maximum, asymmetric decay, and far
tail.

Over 50--300 eV, the Geant4 time spectra are well described by a mixture
of Gamma functions. A single W1 component is insufficient; W2 reproduces
the main asymmetric shape. Adding a third G3 component to the fixed
W2\_QUAD model reduces the NLL and improves the description of the tail
in all six energy intervals. The energy evolution of the main shape is
compactly described by the W2\_QUAD model.

In both RADEX1 and RADEX2, a small fraction of neutrons have very large
delays and are described by the G3 component. Its mean dimensionless
delay \ensuremath{\langle}x\ensuremath{\rangle} is 30.7 for RADEX2 and 38.3 for RADEX1. The fraction of such
events is small: 0.034--0.053\% for RADEX2 and 0.0054--0.0148\% for
RADEX1. At large delays, however, the contribution of the two main
components rapidly decreases, so G3 determines the shape of the far
tail. The two targets contain the same total tungsten and water
thicknesses but have different layer arrangements. In all six energy
intervals, the point estimates of the G3 weight are higher for the
multilayer RADEX2 target, which suggests that the far tail of the time
distribution is sensitive to the internal target geometry.

The main part of the distribution also changes with energy and is
described substantially better by the energy-dependent W2\_QUAD model
than by a single constant W2 form. Comparison with the calculated n\_TOF
Tr108 function shows that the three-component Gamma model better
describes its tail region. The parameters of the slow component,
however, differ substantially from the RADEX parameters and must
therefore be determined independently for n\_TOF.

The time-resolution function should be regarded as an independent
physical object that must be validated before resonance-parameter
reconstruction. Uncertainty in its shape, especially in the far-tail
region, is a potential source of systematic error in the determination
of fundamental neutron-resonance parameters.

\appendix
\setcounter{table}{0}\renewcommand{\thetable}{A\arabic{table}}
\section{Coefficients of the RADEX2 W2\_QUAD model}

For u=(E\ensuremath{-}175 eV)/125 eV, second-order polynomials are used:
ln{[}n\textsubscript{i}(E)+1{]}=a\textsubscript{i}\textsubscript{0}+a\textsubscript{i}\textsubscript{1}u+a\textsubscript{i}\textsubscript{2}u\textsuperscript{2}, ln k\textsubscript{i}(E)=b\textsubscript{i}\textsubscript{0}+b\textsubscript{i}\textsubscript{1}u+b\textsubscript{i}\textsubscript{2}u\textsuperscript{2}, logit
\ensuremath{\epsilon}\textsubscript{2}(E)=c\textsubscript{0}+c\textsubscript{1}u+c\textsubscript{2}u\textsuperscript{2}. The time scale is t\textsubscript{s}(E)=0.5/\ensuremath{\sqrt{E}} \ensuremath{\mu}s (E in eV). The
listed coefficients correspond to the final RADEX2 fit using the time
variable T over 50--300 eV (N=1 653 817).

\begin{longtable}{@{}
  >{\raggedright\arraybackslash}p{(\columnwidth - 6\tabcolsep) * \real{0.2500}}
  >{\raggedright\arraybackslash}p{(\columnwidth - 6\tabcolsep) * \real{0.2500}}
  >{\raggedright\arraybackslash}p{(\columnwidth - 6\tabcolsep) * \real{0.2500}}
  >{\raggedright\arraybackslash}p{(\columnwidth - 6\tabcolsep) * \real{0.2500}}@{}}
\caption{Coefficients of the final RADEX2 W2\_QUAD model.}\label{tab:A1}\\

\toprule\noalign{}
\begin{minipage}[b]{\linewidth}\raggedright
Parameter
\end{minipage} & \begin{minipage}[b]{\linewidth}\raggedright
\textbf{const}
\end{minipage} & \begin{minipage}[b]{\linewidth}\raggedright
\textbf{u}
\end{minipage} & \begin{minipage}[b]{\linewidth}\raggedright
\textbf{u\textsuperscript{2}}
\end{minipage} \\
\midrule\noalign{}
\endhead
\bottomrule\noalign{}
\endlastfoot
ln(n\textsubscript{1}+1) & 1.50355308964 & 0.0622226235557 & \ensuremath{-}0.0371567789236 \\
ln k\textsubscript{1} & 0.202085229501 & 0.0102312856664 & \ensuremath{-}0.0311446997735 \\
ln(n\textsubscript{2}+1) & 1.07845670996 & 0.0446706200449 & \ensuremath{-}0.122476568873 \\
logit \ensuremath{\epsilon}\textsubscript{2} & \ensuremath{-}1.7521959658 & 0.156387506699 & \ensuremath{-}0.197057521819 \\
ln k\textsubscript{2} & \ensuremath{-}0.977522003687 & \ensuremath{-}0.0245761248068 & \ensuremath{-}0.159171354312 \\
\end{longtable}

To test the far tail, the main two-component W2\_QUAD model was fixed
and supplemented by a third Gamma component. The shape of this component
was determined jointly over the full investigated energy range, whereas
its relative contribution was estimated separately in each energy
interval. The fit required only that the third component correspond to
longer delays than the main slow W2\_QUAD component; its position was
not fixed in advance. This approach tests for the presence of an
additional time scale without changing the already determined main shape
of the resolution function.

\section*{Acknowledgments}
The author gratefully acknowledges I. I. Tkachev, G. K. Matushko, and D.
V. Hlustin for their support and assistance.

\section*{Declaration of generative AI use}
ChatGPT Work (OpenAI), primarily using the GPT-5.6 Sol Light model, was
employed as an assistive tool for discussion of the analysis
methodology, development and verification of computational scripts,
organization of the manuscript, translation, and language editing. All
scientific decisions, calculations, interpretation of the results,
references, and the final manuscript were independently reviewed and
verified by the author, who assumes full responsibility for the
accuracy, originality, and integrity of the work.

\section*{References}
\begin{quote}
{[}1{]} S. Agostinelli et al., "GEANT4---a simulation toolkit," Nucl.
Instrum. Methods Phys. Res. A, vol. 506, no. 3, pp. 250--303, 2003, DOI:
10.1016/S0168-9002(03)01368-8.

{[}2{]} R. M. Djilkibaev and D. V. Hlustin, ``INES Time-of-Flight
Spectrometer,'' Instruments and Experimental Techniques 68(2), 226--236
(2025). DOI: 10.1134/S0020441225700356.

{[}3{]} F. Gunsing, ``Generalities on the Time-of-Flight Resolution
Function,'' Consultants' Meeting on EXFOR Data in the Resonance
Region and Spectrometer Response Function, IAEA, Vienna, 8--10 October
2013; INDC(NDS)-0647 (2013).

{[}4{]} C. Guerrero et al., ``Performance of the neutron
time-of-flight facility n\_TOF at CERN,'' Eur. Phys. J. A 49, 27 (2013).
DOI: 10.1140/epja/i2013-13027-6.

{[}5{]} N. M. Larson, Updated Users' Guide for SAMMY: Multilevel
R-Matrix Fits to Neutron Data Using Bayes' Equations,
ORNL/TM-9179/R8, ENDF-364/R2, Oak Ridge National Laboratory (2008).

{[}6{]} D. P. Barry et al., \mbox{``\textsuperscript{2}\textsuperscript{3}\textsuperscript{6}}U resonance parameters at 5.467 eV
from neutron transmission measurements using thin liquid samples,''
Progress in Nuclear Energy 86, 11--17 (2016). DOI:
10.1016/j.pnucene.2015.10.001.

{[}7{]} P. \v{Z}ugec et al., ``Machine learning based parametrization
of the resolution function for the first experimental area of the n\_TOF
facility at CERN,'' Nuclear Science and Techniques 36, 235 (2025). DOI:
10.1007/s41365-025-01820-2.

{[}8{]} D. A. Brown et al., ``ENDF/B-VIII.0: The 8th Major Release of
the Nuclear Reaction Data Library with CIELO-project Cross Sections, New
Standards and Thermal Scattering Data,'' Nuclear Data Sheets 148, 1--142
(2018). DOI: 10.1016/j.nds.2018.02.001.

{[}9{]} M. F. Crouch, ``Experimental Measurement of the
Slowing-Down Time Distribution for Neutrons in Water,'' Nuclear Science
and Engineering 2, 631--639 (1957). DOI: 10.13182/NSE57-A25430.

{[}10{]} A. K. Ghatak and T. J. Krieger, ``Neutron Slowing-Down
Times and Chemical Binding in Water,'' Nuclear Science and Engineering
21, 304--311 (1965). DOI: 10.13182/NSE65-A20033.
\end{quote}

\end{document}